\documentclass[10pt]{article}

\usepackage{latexsym}
\usepackage{verbatim}
\usepackage{graphicx}
\usepackage{mathptmx}
\usepackage{float}
\usepackage{changepage}
\usepackage[normalem]{ulem}
\usepackage{wrapfig}
\usepackage[margin=1.5in]{geometry}
\usepackage{csquotes}
\usepackage{caption}
\usepackage{booktabs}
\usepackage{afterpage}

\fontfamily{ptm}

\title{Belief places and spaces: Mapping cognitive environments}
\author{Philip Feldman\\
	feld1@umbc.edu
	\and
	Aaron Dant \\
	aaron.dant@asrcfederal.com
	\and
	Wayne Lutters \\
	lutters@umd.edu
}
\begin{document}
	
	\maketitle
	
	\begin{abstract}
		Beliefs are not facts, but they are \textit{factive} -- they \textit{feel} like facts. This property is what can make misinformation dangerous. Being able to deliberately navigate through a landscape of often conflicting factive statements is difficult when there is no way to show the relationships between them without incorporating the information in linear, narrative forms. In this paper, we present a mechanism to produce maps of \textit{belief places}, where populations agree on salient features of fictional environments, and \textit{belief spaces}, where subgroups have related but distinct perspectives. Using a model developed using agent-based simulation, we show that by observing the repeated behaviors of human participants in the same social context, it is possible to build maps that show the shared narrative environment overlaid with traces that show unique, individual or subgroup perspectives. Our contribution is a proof-of-concept system, based on the affordances of fantasy tabletop role-playing games, which support multiple groups interacting with the same \enquote{dungeon} in a controlled, online environment. The techniques used in this process are mathematically straightforward, and should be generalizable to auto-generating larger-scale maps of belief spaces from other corpora, such as discussions on social media.
	\end{abstract}

	\section{Introduction}
	As humans, we share experiences primarily though stories. As our communication networks connect us to more and more stories, it is becoming difficult to discern credible and trustworthy sources, while avoiding dangerous misinformation from bad or misguided actors. Social mechanisms that exposed charlatans and cheats at small scales and low bandwidths are no match for fiber optics and state-sponsored troll farms. As users of these systems, we have no way to see how these strands of interconnect. Paradoxically, unless we already know how to search beyond our own bounds of ignorance, we cannot find these more distant horizons. Modern information retrieval technologies exacerbate this problem. Social networks and search engines tend to emphasize  the stories that make us feel comfortable and align with our biases~\cite{flaxman2016filter}. But these stories can pull us blindly along towards disaster, unaware that there is anything outside the bounds of our own constrained beliefs. 
	
	Stories are linear constructs, and are more naturally suited to the presentation os a singular point-of-view over time. We know that Odysseus spent years wandering the Mediterranean region, but only know the \textit{order} of his encounters.  The \textit{Odyssey} though rich in detail, does not provide the spatial relationships that could let another ship avoid the island of the Cyclops~\cite{finley1978homer}. A mariner of Homer's time would rely on a set of these types of stories, collected over a lifetime, to effectively navigate even small regions. After all, as much as it is about the wrath of gods and the foibles of men, the \textit{Odyssey} is about \textit{being lost}.
	
	Even so, stories have been used historically as the basis of navigation for millennia. Before the 16th century, ship's pilots collected \enquote{navigation stories} into a \textit{rutter} or \textit{pilot book}, that described coastal and open ocean routes in narrative form. Because it is difficult to have explicit spatial relationship between stories, rutters \enquote{exhibit an understanding of physical space as delimited rather than panoramic}~\cite{goldie2015early}. To obtain this \textit{panoramic} view, one needs maps. 
	
	Even if there were no such thing as \textit{objective}, surveyed maps, it could still be possible to build panoramic maps based on a careful synthesis of a large set of personal, subjective descriptions contained in ship's rutters. These narrative \enquote{threads} could be knitted together into a tapestry of narrative that could portray the planet's spatial relationships, based on this collection of individual, unrelated paths. These maps could also contain social context. Interesting places would have more text, boring ones less. Dangerous straights might contain a safer passage that could be distinguished from other, riskier routes. Though these maps would not have the representational rigor that objective maps have, maps based on such \textit{subjective} data could still support navigation between the physical places of the world, while also exposing the belief spaces that our subjective experiences are embedded in.
	
	Can these sorts of maps be developed using the vast oceans of online data we now have at our fingertips? To find out, we need to start out with an understandable, well-characterized environment. Starting with readily available online information from such sources as Twitter, Reddit and even the Wikipedia provide an extremely large and complicated information environment,  full of bots, implicit bias and manipulation. To examine this possibility, we need repeatable, testable environments that we can record our narrative journeys through.
	
	Games represent unique, often bespoke universes - some small and intimate, others large and expansive. Some of these games are simple and linear, such as children's board games where tokens are moved along a single path. Others are tightly constrained by rules, but still provide a combinatorial explosion of possibilities, such as chess. In some, the physical environment world plays a dominant role, for example in bicycle racing. In others, there is nothing but verbal exchanges between players such as in debates. 
	
	For this effort we rely on the remarkable world of fantasy tabletop role-playing-games (FTRPGs), as exemplified by Dungeons and Dragons (D\&D)~\cite{gygax1974dungeons}. Classic roleplaying games are a type of shared storytelling which requires intricate coordination between the participants. With each objective or challenge encountered the group needs to discuss the proper way to proceed forward and unfold the story. These constructed \enquote{play spaces} have many properties of physical environments. Participants travel a co-created narrative trajectory. This process has a form of inertia - it must evolve at a rate that is enjoyable~\cite{martindale1992clockwork}. Frequent, abrupt changes in plot direction are avoided. 
	
	These games are collaborative. Each player knows that they must find a level of consensus that allows the story to continue. Too little consensus, and the party may split up, making an impossibly unwieldy story for the Dungeon Master (DM) to sustain and thus end the game. Too much consensus, and the game may become boring. Ideally, the players bring different perspectives and abilities embodied in the characters (PCs) they play, ranging from wizard elves to fighting dwarves. For each challenge presented by the DM, the players explore the problem and potential solutions, aligning around an approach that is dynamically adjusted during gameplay, often in response to the vagaries of chance as expressed through the roll of the dice. The DM provides the narrative frame for the adventure, and will guide the discussion of the players, suggesting solutions and compromises. 
	
	Conversations of this type that need to come to consensus are a unique type of group problem solving. They depend more on alignment than compromise. Moscovici, in  \textit{Conflict and Consensus} shows that \enquote{consensual participation probably has the effect of raising the level of collective involvement} and that this process includes a form of alignment that leads a group along a path of action~\cite{moscovici1994conflict}.
	
	This process of alignment represents movement along an axis through a shared belief space, which consists of the following steps:
	
	\begin{itemize}
		\item The group determines what it is going to discuss. Problems can be approached in many ways, and individuals will often have a unique perspective. To arrive at consensus, the group has to implicitly agree on the salient points they will be debating. This process takes the high-dimensional space associated with each member's perspective and collapses it to the much lower number of arguable points - often simply pro/con.
		
		\item The group can align in the direction of one of the poles. In this case the opinions of all the group's members will move in that direction. Once established, this process is dynamic, and can continue towards more extreme beliefs as the group self-reinforces. Moscovici calls this process polarization, but due to the political overloading of the term, we refer to this process as \enquote{alignment} throughout this paper.
		
		\item Alternatively, the group can work out a compromise, determining the average position. Compromises represent a an intersection of beliefs that all parties can accept, but do not represent any particular alignment. As such these are less dynamic, since they represent a small area of overlap in the parties belief spaces.
	\end{itemize} 
	
	In the world of D\&D, these decision spaces are constrained by the game mechanics. Each room contains a challenge that the party must overcome, usually in the form of an adversary or a puzzle. A sequence of these challenges, organized by the DM, represents an adventure. These adventures can be repeated for multiple parties, and there are numerous guides for sale that contain fully fleshed out frameworks~\cite{ewalt2014dice}. In a FTRPG, there are the agreed-upon \textit{places} in a dungeon -- such as a room with a pit and the belief \textit{spaces} that involve what to do in these places. Belief places should be common to all groups that travel through a particular dungeon, and should be embodied in the artifacts used to create the dungeon. Notes, pictures and books that the DM would refer to. The conversations and the decisions that each group makes in these imaginary places are the cognitive spaces that are latent in the room's configuration. If there is a troll in the room, then the discussions will be centered around the troll, but they will also reflect the unique ways that the challenge of the troll can be overcome. 

	Can a map be built from these texts are built from the interactions between players and the universe they are co-creating? The \enquote{ground truth} in these places (e.g., which room is where) is known to the DM, but must be inferred by the players. The collaborative construction of the multidimensional belief spaces that represent the solution to each dilemma is also adjacent to the physical places. Can these also be extracted automatically to produce maps that are meaningful to humans? In this paper, we describe how we can use the affordances of interactive online text combined with the structures of D\&D that allow us to build proof-of concept maps that demonstrate this possibility.

	\section{Literature Review}
	This work depends on research in multiple fields. Agent-based simulation provides an effective mechanism for representing testable, extensible models of group cognitive interaction. To represent these interactions, we use cartographic affordances and understandings in a human-centered way. Lastly, we employ the mechanisms and culture of tabletop role-playing games, a vigorous and long-lived community that has embraced computer-mediated systems for collaborative storytelling.

	\subsection{Agent-based simulation}
	Animal models have often served as a starting point for understanding human interaction with information. Danchen et. al. showed that animals and humans both use inadvertent social information to influence decisions about environmental quality and appropriateness~\cite{danchin2004public}. Card and Pirolli~\cite{pirolli1999information} successfully demonstrated  the utility of animal models for individual human information foraging behaviors. Deneubourg and Goss'~\cite{deneubourg1989collective} work related to animal group cognitive behaviors such as flocks and herds. More recently, Olfati--Saber et. al. have shown that social influence leading to collective behaviors is a widespread phenomenon in natural and artificial systems~\cite{olfati2007consensus}. Reynolds~\cite{reynolds1987flocks}, Cucker~\cite{cucker2007emergent}, Olfati--Saber~\cite{olfati2006flocking} and others have built and described agent-based simulations that produce emergent flocking, schooling, and herding characteristics that closely mimic observed animal behavior. Connecting animal models to technology-mediated human group interaction, Belz et. al. have shown the emergence of spontaneous flocking in computer mediated communication~\cite{belz2013spontaneous}.
	
	The scientific study of human mass behavior has roots in the 19th-century work of LeBon~\cite{le1897crowd}, who showed that crowds can move and think like single organisms. The mechanisms for these collective behavior, including consensus, compromise, polarization and extremism were studied experimentally by Moscovici~\cite{moscovici1994conflict}.  More recently, Krause~\cite{hegselmann2002opinion} and Bikhchandani~\cite{bikhchandani1992theory} have modeled opinion dynamics and echo chambers while Salganik~\cite{salganik2008leading} has demonstrated that online rating based on popularity can produce runaway results. Epstien et. al shows similar results for Information retrieval with search engines~\cite{epstein2015search}.
	
	\subsection{Cartography}
	Over the past few years, the idea of using embeddings to represent terms, topics and documents in a spatial context has gained currency, particularly in the machine learning community~\cite{collobert2011natural}. The idea of using these types of techniques to produce maps of human knowledge is seductive, but misguided in its current form. This is because although these systems can create a spatial relationship between items in an embedding space~\cite{yao2018dynamic}, it is not \textit{navigable} in a \textit{human} context. The way that machines create these spaces does not sufficiently take into account the manifold relationships that humans have with spatial and symbolic data. 
	
	Maps are a special subset of diagram, that contain a spatial and symbolic aspect such that it is possible to understand the geometric relations between elements~\cite{fathulla2007diagram}. In a traditional map, and example of this difference would be political borders. The geographic (spatial) representations of the terrain features are present in the world, but the symbolic borders are a manifestation of beliefs that we as humans bring to the environment, and do not exist in physical space and yet have locations. 
	
	Because we can \enquote{see ourselves in the map}, we can extrapolate from the places we're been to places that we've never seen. Maps can inform us of what we are likely to encounter on our journey. Building the maps \enquote{gives materiality and objectivity to space}~\cite{jacob2006sovereign}. Finally, because of this shared representation that maps can share at a glance, diagrams such as maps provide a type of communication that is distinct from other forms such as lists and stories. They afford reasoning about time and space~\cite{isozaki1992mechanism}, even if the locations are imaginary~\cite{rohl2008mapping}. 
	
	\subsection{Role-playing games}
	Using role-playing games (RPGs) as a bridge between full simulations and \enquote{unstructured human interaction in the observed world} has been tested by~\cite{barreteau2001role}, who compared simulation and RPGs of irrigation practices in the Senegal River valley. RPGs have also been used for decades to co-create interactive fictions that can explore widely varying issues and settings. These FTRPGs were developed by~\cite{gygax1974dungeons}. They took existing tabletop wargaming affordances and changed them so that small groups of individuals could play as characters within a semi-structured environment. FTRPGs are systems that take a latent structure (using commercially-available references or created by the DM), and, through the process of gameplay, build emergent structures of specific actions based on the interactions of the players and their evolving alignment on how to proceed through the current scenario~\cite{borgstrom2005structure}. This combination of constrained options that afford free-form responses provides a framework whereby multiple groups can come to consensus about many different kinds of problems, ranging from the practical (how do we get across the pit) to the moral (do we kill the guard?). These vast narratives are capable of supporting intimate, honest discussion between the protagonist players, both in and out of character~\cite{harrigan2010second}. The terrain that these narrative structures become Foucault's \textit{other spaces}, and exist as alternate realities in the player's minds~\cite{foucault1986other}.
	
	\subsection{Group cognition}
	Why should we expect that groups interacting in these spaces should produce the kind of data that we need to produce maps of shared cognitive spaces? The answer lies in the dimension-reducing process that groups use to achieve alignment. Moscovici showed that this process of alignment has several steps. Recall that the group has to implicitly agree on what they are going to discuss. All the individuals with their own, unique perspectives have to create a common, shared reality that they can then debate. Uncommon perspectives are literally incomprehensible on a neurological level to people who have not encountered them previously~\cite{yeshurun2017same, parkinson2018similar}. Once the group has arrived at this lower dimensional space, they can either compromise and average the beliefs of the group, or orient along an axis that affords argumentation with respect to the poles of the discussion. This process of alignment moves the average belief of the group towards one of the poles~\cite{moscovici1994conflict}. It is this process of dimension reduction and alignment that makes us believe that it is possible to build low-dimension, human-comprehensible maps.

	\section{Our Model}
	\begin{figure} [h]
		\centering
		\fbox{\includegraphics[width=0.9\columnwidth]{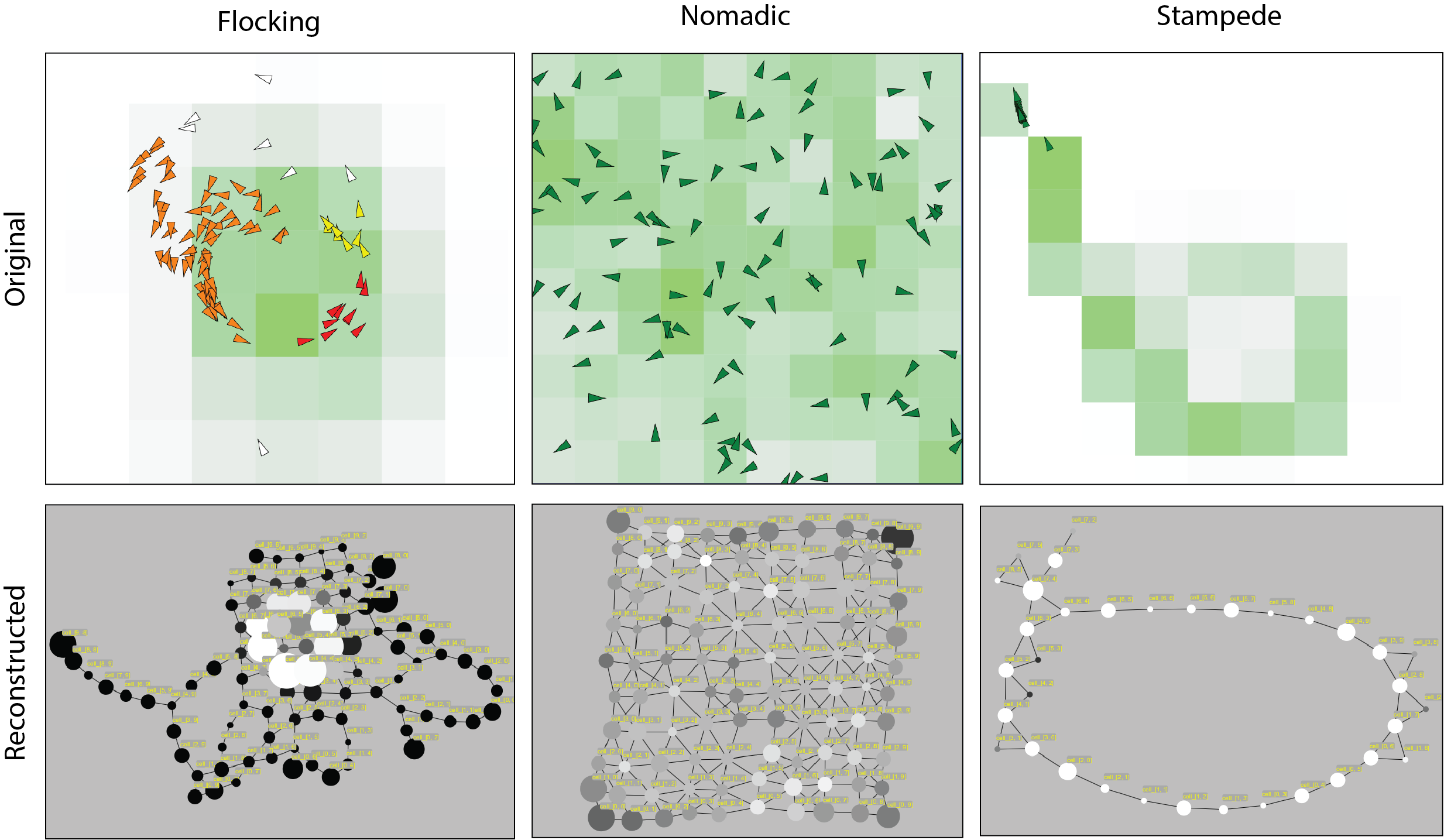}}
		\caption{Simulations (top) and associated maps (bottom)}~\label{fig:sim_map}
	\end{figure}
	
	We have led a multi-year effort to produce maps of belief space using human-generated content. Our work takes inspiration from the parable of Simon's Ant~\cite{simon1996sciences}, where an ant follows a complex path along a beach on to reach its goal. The story shows how the interaction of an agent incorporating simple rules and a complex environment can produce complex behavior. But this relationship is commutative -- given the rules and enough ant paths, we can infer the environment. With a large population of agents -- human or animal -- it should be possible to build a detailed map of a population's social and physical environment by examining and aggregating the information seeking behavior of individuals~\cite{feldman2018one}. 
	
	We began with agent-based simulations to develop theoretical models of population-level behavior in belief space. Since people can have beliefs along many axis, agents exist in an n-dimensional hypercube, and interact according to a modified \enquote{boids} algorithm~\cite{reynolds1987flocks}. Agent behavior is affected by the social influence horizon (SIH), in which influence falls off as a function of distance in the belief space coordinate frame. A very low SIH means that agents are influenced only by nearby agents and are unaffected by other agents with more distant beliefs. A high SIH means that all agents influence each other equally, regardless of distance. Fewer dimensions effectively move individuals closer together, and amplify SIH effects. This agrees with Moscovici's finding that for groups to align, they have to reduce the dimensions of the discussion to support alignment~()\cite{moscovici1994conflict}).
	
	The effects of varying the SIH can be viewed as creating networks that evolve over time. For low SIH, nomadic agents are connected to few or no other individuals, resulting in behavior that emerges solely from the agent's interaction with the environment. Moderate SIH connects clusters of agents together in small networks, while an infinite SIH connects all agents equally. 
	
	Behaviors that emerge by varying the SIH resemble models of social animals moving socially in physical space. This can be seen in the top row of figure \ref{fig:sim_map}.  A moderate distance leads to the emergence of ad-hoc \textit{flocks} that tend to cluster in the center of the hypercube. Densely connecting all agents with a high SIH causes all agents to align and collapse into a single \textit{stampede}. The level of social interaction is inversely proportional to the number of dimensions. 
	
	Once we had reliable models of behavior in belief space, we worked on mechanisms that could record agent semantic behavior that resemble online textual interaction, such as social media posts. Each element of the hypercube has one or more text \enquote{statements} that can be read and \enquote{posted} by an agent traveling through it. This process creates a trajectory of textual strings that resemble, for example, and evolving set of statements that start at one \enquote{position} and evolve to a different perspective. 
	
	The published history of each agent is a textual trajectory that can be used to reconstruct a representation of the environment that contains physical and social manifestations~\cite{feldman2018maps}. A can be seen in the bottom of figure \ref{fig:sim_map}, the nomadic agents produce a map that most accurately reflects the shape of the underlying environment (in this case, a 2-D square). The flocking agents create an accurate reconstruction of the center of the environment, while the stampeding population has very little representation of the underlying space, but shows that the agents have behaved as a highly unified group.

	The current phase in this effort is to validate these models using human interactions in a well-characterized, bounded, environment. Using the semi-structured play affordances of FTRPGs, we will show that it is possible to build small-scale maps from human online textual interaction. The goal is to create automated processes that can be used to build a \enquote{map factory} that scales to large-scale belief environments, providing a mechanism that can be used to let users locate themselves and navigate across belief environments.

	\section{Study Design}
	\label{sec:study_design}
	Although \enquote{the map is not the territory}~\cite{korzybski1958science}, one of the unique features of a map is the ability of readers to see themselves in it -- to recognize their surroundings. For a map to be useful as a map and not just a diagram, relationships between objects need to be portrayed in such a way that it is possible to \textit{plot a course} that accurately predicts what the traveler will encounter.  Lastly, a map needs to be able to support communication. One user should be able to communicate meaningful information about a course that another might be ready to embark on. As the goal of this research is to produce \textit{human-usable} maps, we needed to observe and record human activity in a semi-structured environment. This environment needed to be simple enough that we could develop baseline techniques that could be shown to work against a known environment (ground truth), while also being able to identify aspects of group behavior that are not directly tied to the environment.
	
	\subsection{Game environment considerations}
	To build our fictional environments, we used the well-known D\&D 5th edition rules~\cite{heinsoo2008dungeons}. For this proof-of-concept method and practice, we constructed a single dungeon designed to be simple and consistent. Multiple parties would be run through this environment, and the text of their interaction s would be recorded and stored. This initial environment consisted of a predefined play area (the dungeon) with obstacles and challenges to overcome. The dungeon was linear and sequential, consisting of four rooms connected by magical gates that would close after the party members passed through them. A framing story and room descriptions were identically presented to each party using of pre-written introductions. The Dungeon Master (DM) pasted these texts into the text stream at the beginning of the adventure, the beginning of each room, and at the adventure's conclusion. An example snippet from the introduction to the first room is shown below:
	
	\hspace{1cm}\begin{minipage}{\dimexpr\textwidth-2cm}
		\textit{\enquote{Peering through the vines, you can see the antechamber opens to a platform with circular stairs descending to the left and right. Beyond the platform the room is a large circular chamber with a domed roof mostly lost to shadows. Scraggly trees and grass have grown up through the stones of the floor making it seem as if a natural clearing in ancient ruins lies below you on the floor of the next room. The light revealing all of this comes from a single moderately sized campfire assembled in a circle of broken stones. The flickering firelight reflects on what appears to be a golden gate on the far side of the room.}}
	\end{minipage}
	
	An adventure party consisted of four players. Each group of players were given a limited set of pre-generated characters to choose from. Once characters were chosen and named by the players, the group began the game, encountering the same scenarios in the same order as every other group. Where positioning was an important factor, a floorplan was provided. Players were presented with a linear sequence of four rooms, since that was determined to be complicated enough to develop the text analytics against, but also simple enough so that reasonable data could be gathered with relatively short runs, and with a smaller sample size (Five groups)
	
	Players were not told anything about the dungeon beforehand, and were strongly encouraged not to talk to other potential players. DMs that were not one of the authors were selected from players who had completed the adventure. The considerations involved in constructing the game environment for this research effort is discussed in detail in~\cite{dant2019dungeons}.
	
	\subsection{User interface and interactions}
\begin{figure} [h]
		\centering
		\fbox{\includegraphics[width=0.75\columnwidth]{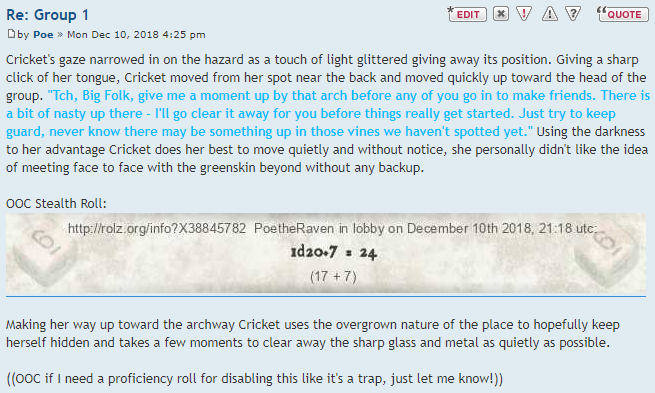}}
		\caption{A play-by-post submission including character actions, speech, dice rolls, and OOC comments in one contiguous block}~\label{fig:PHPBB_post}
	\end{figure}

	Initially, to provide an historically familiar environment for textual interaction, a PHP-based bulletin board (BBS) was instantiated and hosted on a private server. Play-by-post (PBP) is a method of roleplaying has its roots in the original bulletin boards of the 1980s and surprisingly has not changed much in the nearly 40 years since its inception~\cite{Samory2017Quotes}. Posts (character and player) tend to range from a few sentences to paragraphs. An example is shown in figure \ref{fig:PHPBB_post}. Users were encouraged to use selected fonts and colors to differentiate between dialog, action, and out-of-character (OOC) comments, though this information is not currently used in our analytics.
	
\begin{figure} [h]
		\centering
		\fbox{\includegraphics[width=0.75\columnwidth]{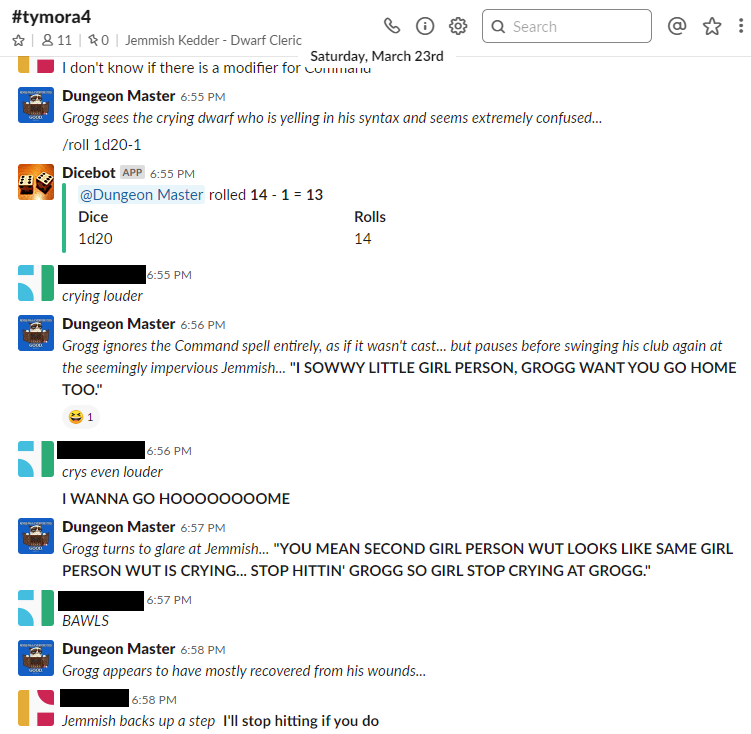}}
		\caption{A real-time chat example with interleaved conversation and dice rolls between multiple participants}~\label{fig:slack_post}
	\end{figure}

	Although the BBS-based system worked very well for the first group, which consisted of experienced players used to the affordances of play-by-post, it turned out to be too difficult for subsequent groups. We thought that using interactive messaging affordances, such as those provided by Slack\footnote{slack.com}, we might achieve better participation results. This assumption turned out to be true, and the remaining four groups were run on this platform. Typically, a group would complete the game in approximately eight hours, usually split across two sessions. The Slack interface produced shorter, more rapid interactions. Posts could be as short as one word and would seldom be longer than two sentences.  An example of this posting style is shown in figure \ref{fig:slack_post}.
	
	Although these two text styles appear very different, the analytics that we used were able to effectively handle these variations, as discussed in section: \textit{\ref{section:results}}.
	
	\subsection{Participant Recruiting}
	A variation of snowball sampling was used, where players were recruited from the online and physical player communities that the primary dungeon master and co-author (A. Dant) was a member. Additional members were recruited from communities that the initial participants were also members. Player experience ranged from novice to extremely experienced players with prior experience as DMs in a variety of FTRPG platforms. Each member provided informed consent. 
	
	\subsection{Analytics and map production} \label{analytic-design}
	Two data sets were produced in this effort, one from each platform. These files were placed in a MySQL database. SQL views were written to make queries identical between the two data sets so they could be treated as a single corpus.
	
\begin{figure} [h]
		\centering
		\fbox{\includegraphics[width=0.75\columnwidth]{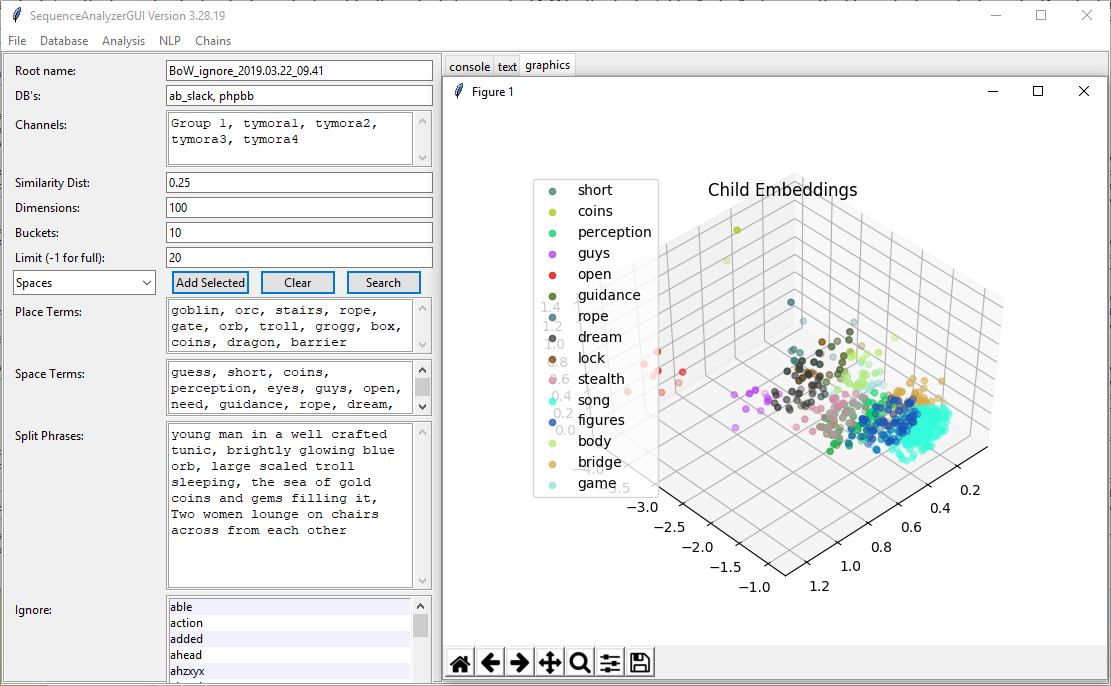}}
		\caption{Data explorer with term embeddings}~\label{fig:tool}
	\end{figure}

	Initial, \textit{ad-hoc} data exploration was slow, complicated, and prone to error. To streamline the workflow, an interactive tool was developed that  automated and simplified the federating of post-related data across the multiple adventures (Shown in figure \ref{fig:tool}). The tool saves and loads human-editable xml configuration files in order to consistently repeat analysis. 
	
	The analytic pipeline was built to support a variety of natural language processing (NLP) approaches. All word data sets could be filtered for stop words, group, individual, and time:
	\begin{enumerate}
		\item Bag of Words (BoW) ~\cite{kosala2000web}: Sequential bags of words (SBoW) could be analyzed along a variety of axis, including individual player, group, group by time slice, all groups by time slice, and all groups
		\item Term Frequency - Inverse Document Frequency (TF-IDF)~\cite{kosala2000web}: The same processes developed for BoW were extended to incorporate TF-IDF \enquote{Documents} in this case were the individual posts when analyzing single players, players in groups, and groups within time slices 
	\end{enumerate}
	
	\subsection{User evaluation} \label{section:user_eval}
	To close the loop on human usability of the maps, we developed a survey for the participants of the dungeon, hosted on Google Forms. This survey was run after all groups had traversed the dungeon, and the initial map based on the full corpora of text interactions was created. The survey showed a picture of the generated map, and asked the following questions. Each question had a positive, neutral, and negative response (\enquote{yes}, \enquote{maybe}, \enquote{no}). The three-point scale was deemed appropriate, due to the coarseness expected in the evaluation of this novel presentation. 
	\begin{enumerate}
		\item \textit{Helpfulness}: If you could have consulted the map during gameplay, Would you consider it? 
		\item \textit{Recognition}: Can you see your experiences in the map?
		\item \textit{Accuracy}: Do the descriptions of the rooms and the surrounding adventure-specific terms seem appropriate?
		\item \textit{Effective communication}: If you we describing the dungeon to someone who had not run it, Would the map help?
	\end{enumerate}
	Additionally, two open-ended questions were asked to provide additional context and framing for the responses~\cite{dillman2014internet}:
	
	\begin{enumerate}
		\item Would you have done anything differently in your adventure if you had had this map?
		\item Comments (Any questions we should have asked?)
	\end{enumerate}

	\section{Results} \label{section:results}
	A total of five full sessions and one partial session were run. One complete session and one partial session were run using a BBS forum based system. The other four sessions were run on Slack. A MySQL database was developed that contained information from 23 players producing 9,709 usable posts.
	
	Analytic software was written in Python to construct queries, read from the database and store term usage statistics as described in section \ref{sec:analysis}. To ensure consistency between sessions, incidental chats before the game began and after the game ended were excluded, meaning that only player posts during gameplay were used. Further, only complete sessions were used. The data was partitioned as follows:
	
	\begin{enumerate}
		\item A record of all terms used by each player.
		\item A record of all terms used by each group of players.
		\item A record of the aligned text for the same environment across all groups. These were calculated by finding marker posts with maximum similarity across all groups, such as when a party encountered a new room.
		\item Timestamps of each sequence start/stop marker posts by each group.
		\item A record of all terms from all users. 
	\end{enumerate}
	
	Although software was written to apply more sophisticated text analytics such as word embeddings and document centrality, initial results simply using sequential bag-of-words (SBoW) proved startlingly effective, and were used for this proof-of-concept effort. An incidental benefit of this approach was that we were able to describe the algorithm to the participants of the study, which improved their credibility judgment of the results. 
	
	This analysis was broken into four sequential stages.
	
	Since we were using a SBoW approach initially, we had to determine stop words. This was done by using the NLTK list of English stop words\footnote{gist.github.com/sebleier/554280}, game-specific terms such as \enquote{d20}, and nonsense strings (e.g. GUIDs) generated by the online systems. Additionally, the disproportionately frequent words from each player when compared to the full corpus from all players in all groups were added to the list. These terms were inevitably terms related to the player's self-identification. This list was saved to a configuration file that was then used for subsequent analysis.
	
\begin{figure} [h]
		\centering
		\fbox{\includegraphics[width=0.7\columnwidth]{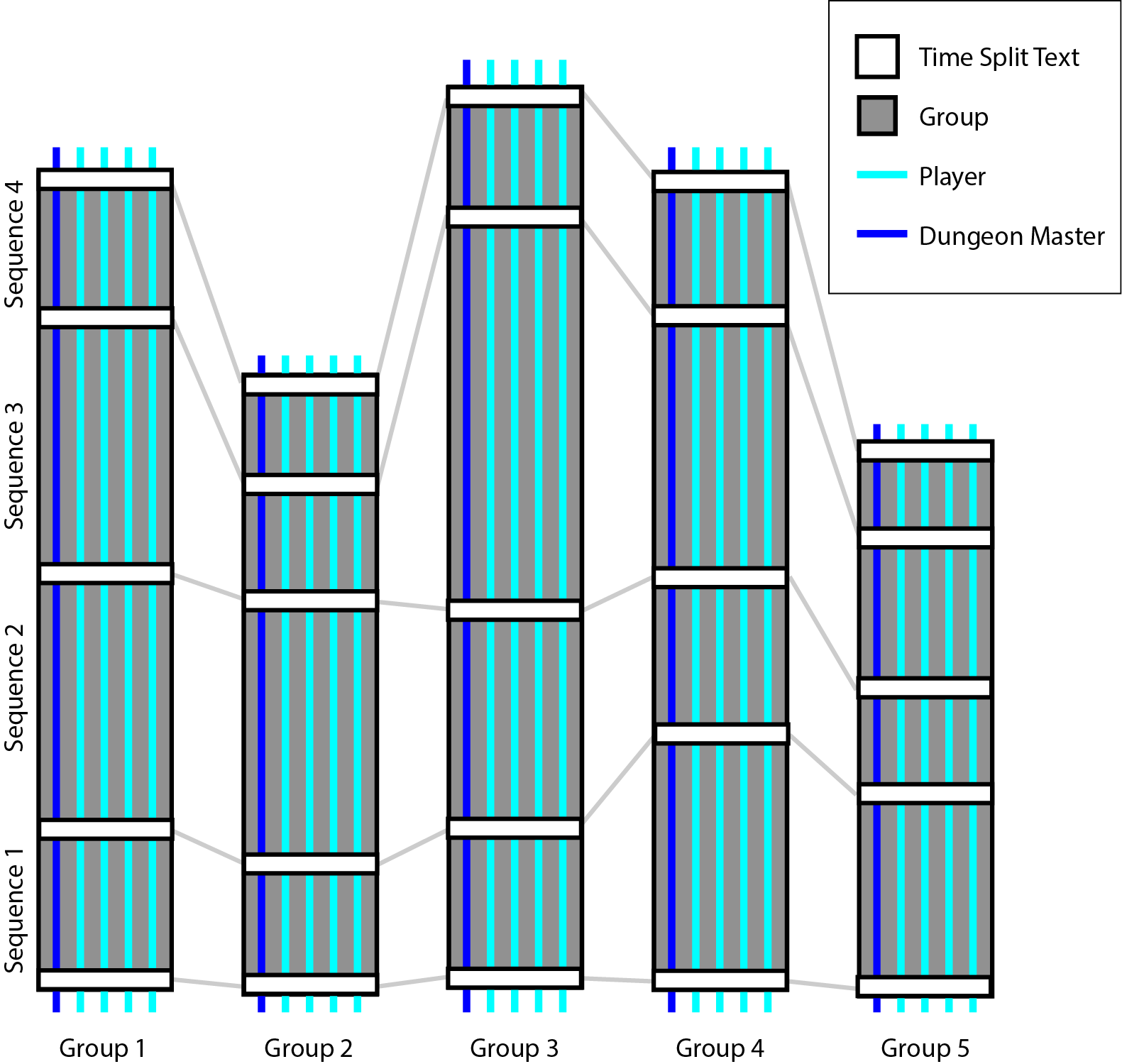}}
		\caption{Text extraction process} \label{fig:extraction}
	\end{figure}

	The next step was to determine the common texts in the narrative. Each group record was compared against the other group records looking for high levels of text matching, such as the room introductions. Between each marker, one or more evenly divided \textit{buckets} of posts could be placed to do more granular word counts. For the dungeon in this paper, there was only one bucket of posts per group per marker, even though the time that elapsed between each marker might range from an hour to several days (Fig \ref{fig:extraction}). Terms were then collected for the same bucket across all the groups and the most common words listed to a depth of 20. This was somewhat arbitrary as can be seen in table~\ref{table:depth_evaluation}, where different depths were tried. For each depth, the top terms reflect an aspect of each group's approach to the challenge. 
	
	\begin{table}[h]
		\hspace{-1.6cm}
		\begin{minipage}[b]{0.6\linewidth} 
			\centering
			\begin{tabular}{rrlllll}
				\toprule
				Depth & Seq & Group 5 & Group 4 & Group 3 & Group 2 & Group 1 \\
				\midrule
				5 & 1 & hit, guys & behind, gate & gate, coins & spell, glass & fire, arrow \\
				10 & 1 & guys, temple & short, log & coins, light & spell, glass & eyes, bow \\	
				20 & 1 & guess, hey & short, stealth & coins, light & percep, antechamber & eyes, side \\
				
				5 & 2 & around, feet & side, pit & side, lever, pit & trap, guys & see, side \\
				10 & 2 & something, percep & tie, wall & lever, trap & trap, guys & anything, metal \\
				20 & 2 & guys, ropes & open, gear & need, dwarves & guys, guidance & anything, halfling \\
				
				5 & 3 & want, see & dream, club & attack, damage & spell, stealth & open, gate \\
				10 & 3 & rope, oil & dream, tiny & attack, damage & spell, stealth & song, sing \\
				20 & 3 & rope, oil & dream, tiny & lock, hit & stealth, guidance & song, sing \\
				
				5 & 4 & light, blue & eyes, body & woman, treasure & woman, stay & grogg, light \\
				10 & 4 & blue, figures & body, enormous & bridge, choose & stay, across & grogg, looking \\	
				20 & 4 & figures, coin & body, offer & bridge, insight & game, boy & grogg, looking \\
				\bottomrule
			\end{tabular}
			
		\end{minipage}
		\caption{Depth evaluation} \label{table:depth_evaluation}
	\end{table}
	
	Top terms for each time slice were then extracted, excluding the 20 terms from the previous step. This was used to show the different actions of each group. The relationship of these passes through the corpora are shown in figure \ref{fig:extraction}. The vertical lines are the players' (including the DM, who plays non-player-characters) individual posts, which are contained in the gray bars that represent each group. The white rectangles are the identified "marker text" used to identify common events in the narrative. Sets of text between these markers are connected by the light gray lines, and represent a common \textit{time slice} or \textit{sequence} across all dungeons that are used to identify the \textit{places}. Each individual sequence within a group is used to determine the belief \textit{space} for that group as it faced the challenges associated with the place. Again, the approach seems resilient with respect to the number of terms chosen. The overall approach of between-groups terms used to label the \textit{places} and within-group terms used to label the \textit{spaces} appears to create sets of terms that make sense to human users.
	
	The most common three elements in each sequential \textit{places} list were used to label each location in the dungeon. These were then connected together to provide the map shown in figure \ref{fig:simple}. Each place node in turn was connected to the most frequent terms used by the groups in that sequence. Once finished, this map was provided to the 23 players who had completed the adventure for evaluation.
	
	Based on feedback from the players, a second map was generated that provided more contextual detail. Each place node was connected in a directed graph to the next node. The place nodes were also connected in a non-directed sub-graph to the most frequent terms used by the groups in that sequence. Lastly, a query was run against the posts to find the shortest post that contained all the terms within the specified time boundaries of the sequence. These posts used to annotate the map and provide context for the terms. For presentation and legibility purposes, the components of the resultant map were manipulated in Adobe Illustrator to see the version shown in figure~\ref{fig:final_map}. We believe that this final step is easily automated, but that task remains for future work.

	\section{Analysis}
	\label{sec:analysis}
	Preliminary analysis was performed once the first run was complete to evaluate text analysis techniques for when data gathering was completed. The text for each member in the group was split into evenly into 5 \enquote{buckets} (splits 1-5) that roughly aligned with the sequence of actions in the dungeon. The top 25\% terms were extracted for each bucket using a combination of BoW/TF-IDF analytics where each players' text in each split was treated as a single document. Once these word lists were created, term centrality~\cite{kritikopoulos2007wordrank} was calculated using the terms from all users within a split. This produced the initial term list shown in table \ref{table:initial_paths}.
	
	\begin{table} [h]
		\hspace{-1em}
		\begin{minipage}[b]{.5\textwidth}
			\centering
			\begin{tabular}{rlll}
				\toprule
				& Group 1 & Group 2 & Group 3 \\ 
				\midrule
				Split 1 & goblin, arrow & stair, spell & behind, goblin \\ 
				Split 2 & statue, halfling & statue, goblin & short, goblin \\ 
				Split 3 & troll, chest & statue, switch & piton, around \\ 
				Split 4 & grogg, troll & troll, sleep & grogg, troll \\ 
				Split 5 & dragon, grogg & grogg, troll & grogg, dream \\ 
				\bottomrule	
			\end{tabular} 
			\caption{Initial, evenly split text} \label{table:initial_paths}
		\end{minipage}%
		\hspace{4em}
		\begin{minipage}[b]{.5\textwidth}
			\centering
			\begin{tabular}{ll}
				\toprule
				& Terms \\
				\midrule
				Sequence 1& goblin-orc-stairs \\ 
				Sequence 2& rope-gate-orb \\ 
				Sequence 3& troll-grogg-box \\ 
				Sequence 4& coins-dragon-barrier \\ 
				\bottomrule	
			\end{tabular} 
			\caption{Final, correlated sequences} \label{table:final_paths}
		\end{minipage} 
	\end{table}

	These terms were then used to produce a connected network using the same technique used to produce the \enquote{Reconstructed} agent maps in figure \ref{fig:sim_map}. This map was shown informally to the participants of the first group to evaluate reactions. Users who had experienced the dungeon were able to see terms that made them feel as though the image captured some aspects of the experience, particularly when the room-specific \textit{places} were highlighted by hand. 

	\afterpage{
		\begin{figure} [h]
			\centering
			\begin{minipage}[t]{0.5\textwidth}
				\centering
				\fbox{\includegraphics[width=0.9\linewidth]{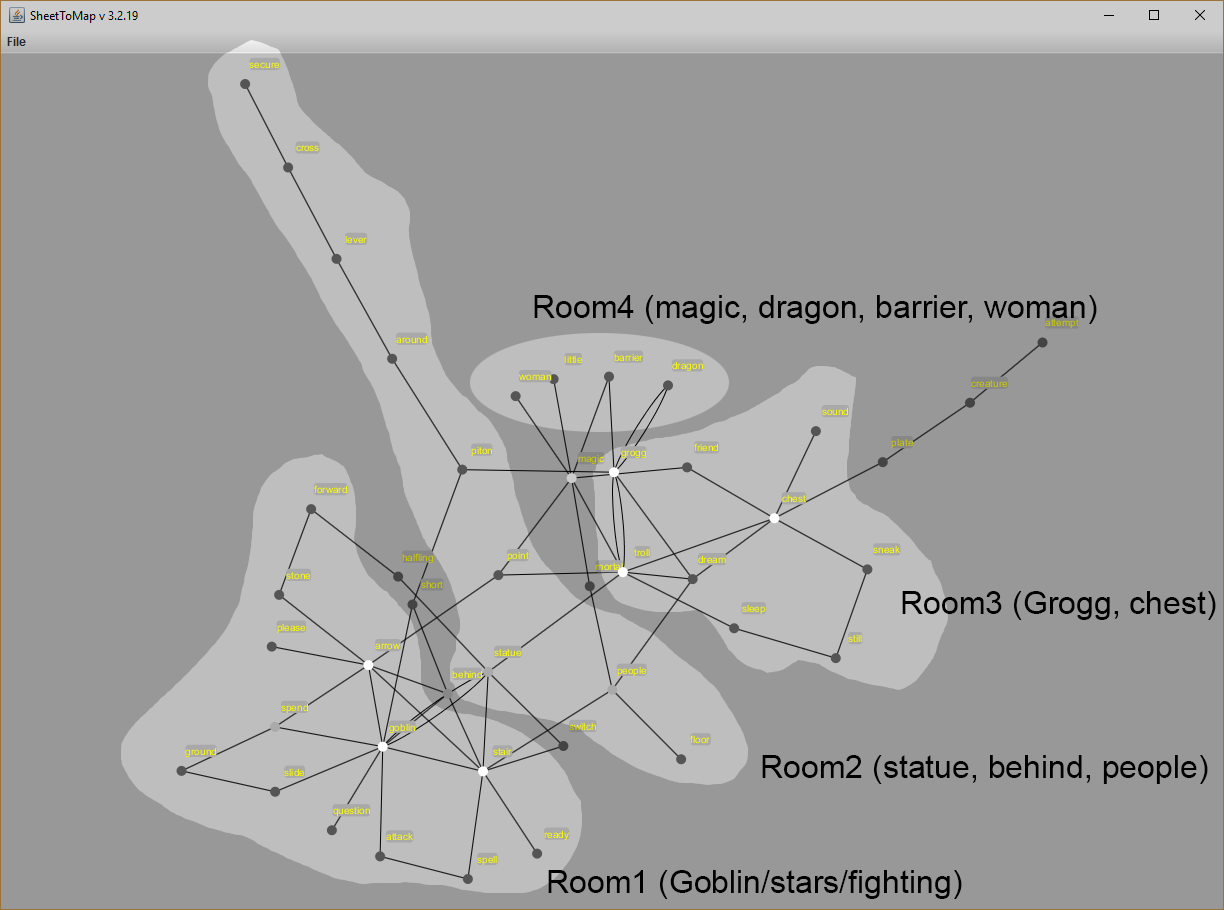}}
				\caption[PoC Map]{Proof-of-concept map\textsuperscript{\footnotemark}} \label{fig:first_map}
			\end{minipage}
			\begin{minipage}[t]{0.45\textwidth}
				\centering
				\fbox{\includegraphics[width=0.9\linewidth]{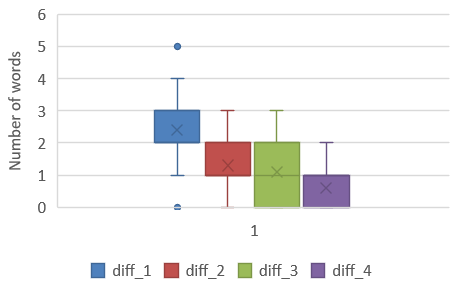}}
				\caption{Convergence of terms}~\label{fig:convergence}
			\end{minipage}
		\end{figure}
		\footnotetext{Full size at antibubbles.com/firstMapAnnotated.png}
	}
	
	Based on this initial output, the need for a more interactive and iterative text analytic (as described in the section \ref{section:results}) became apparent. An interactive tool was built (Figure \ref{fig:tool}) that could rapidly evaluate and diagram different groupings of text. 
	
	Full analysis of all adventures began after five full runs were completed and data from the runs were normalized to account for the different online systems that stored the text. Using the iterative text analytic shown in figure \ref{fig:tool}, we found that the top terms for each sequence across groups stabilized quite quickly. Figure \ref{fig:convergence} shows the convergence on the final terms for all possible permutations and orderings of the groups. In this chart, the whisker plot \enquote{diff\_1} shows the number of terms that vary between any two randomly selected groups. In other words, the 12 terms shown in Table \ref{table:final_paths} could vary by as many as 5 words. By the time all groups are used to create the term list (diff\_4), the maximum difference is 2 terms. It appears that it only takes a few runs through such an environment to extract a common set of terms that can be used to stably label features. After processing, the most common terms for each section are shown in table \ref{table:final_paths}.

	\textit{Belief spaces}, or the points of discussion in each room, were associated with each node by finding the most frequent words for each group after excluding the \textit{places} list of common terms across all groups within the sequence. As with the place terms, the top three terms were chosen. In some cases, as in the \textit{rope-gate-orb} room, the three terms overlapped, implying that teams solved the traps in the room using similar approaches. For example, in several runs, a pit trap opened, that required the use of pitons and ropes. In many cases, but not all, the \textit{halfling rogue} character, who had advantages in disarming traps was involved in opening the gate to the next room without triggering the trap.

	Combined with the time information, an improved diagram of the dungeon \textit {places} was generated, using the Networkx library\footnote{networkx.github.io} (figure \ref{fig:simple}). This version shows the \textit{places} as a connected string of four nodes, with the \textit{spaces} radiating out from their related nodes. This map was provided via email to the study participants in the three-point survey discussed in subsection  \textit{\ref{section:user_eval}}. Of the 23 users, 14 participated in the survey, a response rate of 61\%. The results are shown in table \ref{table:map_one_response}.
	
	\afterpage{
		\begin{figure} [!t]
			\centering
			\fbox{\includegraphics[width=0.85\columnwidth]{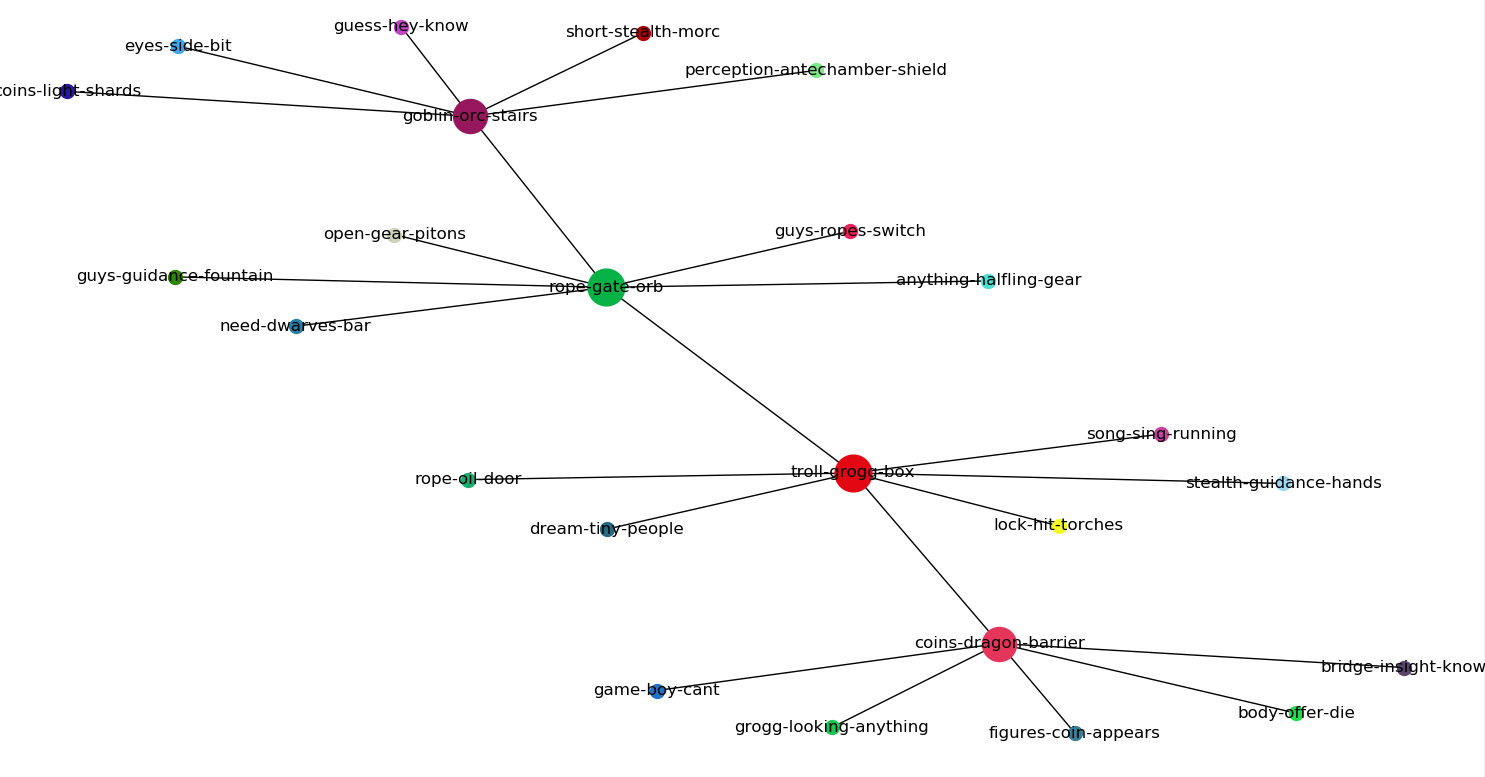}}
			\caption[initial map]{Initial \textit{Belief Places \& Spaces} map\textsuperscript{\footnotemark}} \label{fig:simple}
		\end{figure}
		\footnotetext{Full size at antibubbles.com/simple.png}
	}

	In addition to the 3-point scalar choices, there were two open-ended questions. In response to the \enquote{Would you have done anything differently in your adventure if you had had this map?}, there were a range of opinions, ranging from \enquote{no}, and \enquote{I don't think so}, to \enquote{Walked down the f*\&@ing stairs} (the last quote refers to a particularly dangerous part of the first room). Several of the players were experienced DMs, and had their own perspectives, exemplified by this quote:
	
	\begin{displayquote}
		\textit{\enquote{From past DMing experience, I would have a near certain suspicion that having this map ahead of time would have generated a lot of discussion about the points on the map, focusing the discussion to those points and no consensus would be reached among the players about what each point means.}} (Yanadar: Group 2)
	\end{displayquote}
	
	\begin{table} [h]
		\begin{minipage}{.5\linewidth}
			\centering
			\begin{tabular}{llll}
				\toprule
				\textit{Question}& \textit{No}& \textit{Maybe}&\textit{Yes}  \\ 
				\midrule 
				Helpful?&  26.7\%&  13.3\%&  60\%\\ 
				Recognition?&  &  33.3\%&  66.7\%\\ 
				Accuracy?&  7.1\%&  35.7\%&  57.1\%\\ 
				Effective?&  13.3\%&  46.7\%&  40\%\\ 
				\bottomrule
			\end{tabular} 
			\caption{Map 1 Survey Responses} \label{table:map_one_response}
		\end{minipage}%
		\begin{minipage}{.5\linewidth}
			\centering
			\begin{tabular}{llll}
				\toprule 
				\textit{Question}& \textit{No}& \textit{Maybe}&\textit{Yes}  \\ 
				\midrule 
				Helpful?& 7.1\% &  35.7\%&  57.1\%\\ 
				Recognition?& & 21.4\% & 78.6\% \\ 
				Accuracy?&  & 42.9\% & 57.1\% \\ 
				Effective?& & 28.6\%  & 71.4\% \\ 
				\bottomrule 
			\end{tabular} 
			\caption{Map 2 Survey Responses} \label{table:map_two_response}
		\end{minipage}
	\end{table}
	
	Additional comments centered on a lack of directedness in the map (the rooms were connected by magic doors that would close after passage), and a lack of context - \enquote{The map doesn't give enough information to modify my actions} (Edmund DeVir: Group 2). One user requested an interview, in which we discussed in detail aspects that we could automatically generate based on the content available. Based on this feedback, we created a second map (figure \ref{fig:final_map}). This map now has a directed series of nodes that represent the physical \textit{places} in the dungeon, while each place contains nodes that show the group-specific discussion \textit{space} terms. Posts from the database that contain all three place or space terms (e.g. \enquote{goblin}, \enquote{orc}, \enquote{stairs}) in the correct time slice were retrieved if available and shortened to 160 characters, since some posts could be very long. Lastly, the map was formatted by hand to be legible on a monitor.
	
	A second survey, identical except for the presentation of this new map, was given to the players. Participation was again 61\%. The participants who filled out the form found the new map to be significantly better, as shown in table \ref{table:map_two_response}
	
	As before, the survey had room for open-ended responses. These comments tended to emphasize the value of the snippets as a mechanism for adding context. The following quote shows an example of how the map could help a party of adventurers:
	
	\begin{displayquote}
		\textit{\enquote{Definitely, especially based on the text snippets, and now that I have a better understanding of how to read the map. I would have tried to figure out something using rope much earlier in the orb room since rope seems to be the only commonly used word not directly related to the description of a room or its contents. I would have tried to make friends with Grogg first as it's much clearer from the snippets that he's sentient, talkative, and potentially friendly, and would probably have tried to sing to him if things went badly.}}  (Shelton Harrington: Group 2)
	\end{displayquote}
	
	As these are very early results, we believe that there is considerable ground for further improvement of the map, in particular using more affordances of cartography. 
	
	\afterpage{
		\begin{figure} [!t]
			\centering
			\fbox{\includegraphics[width=1.0\columnwidth]{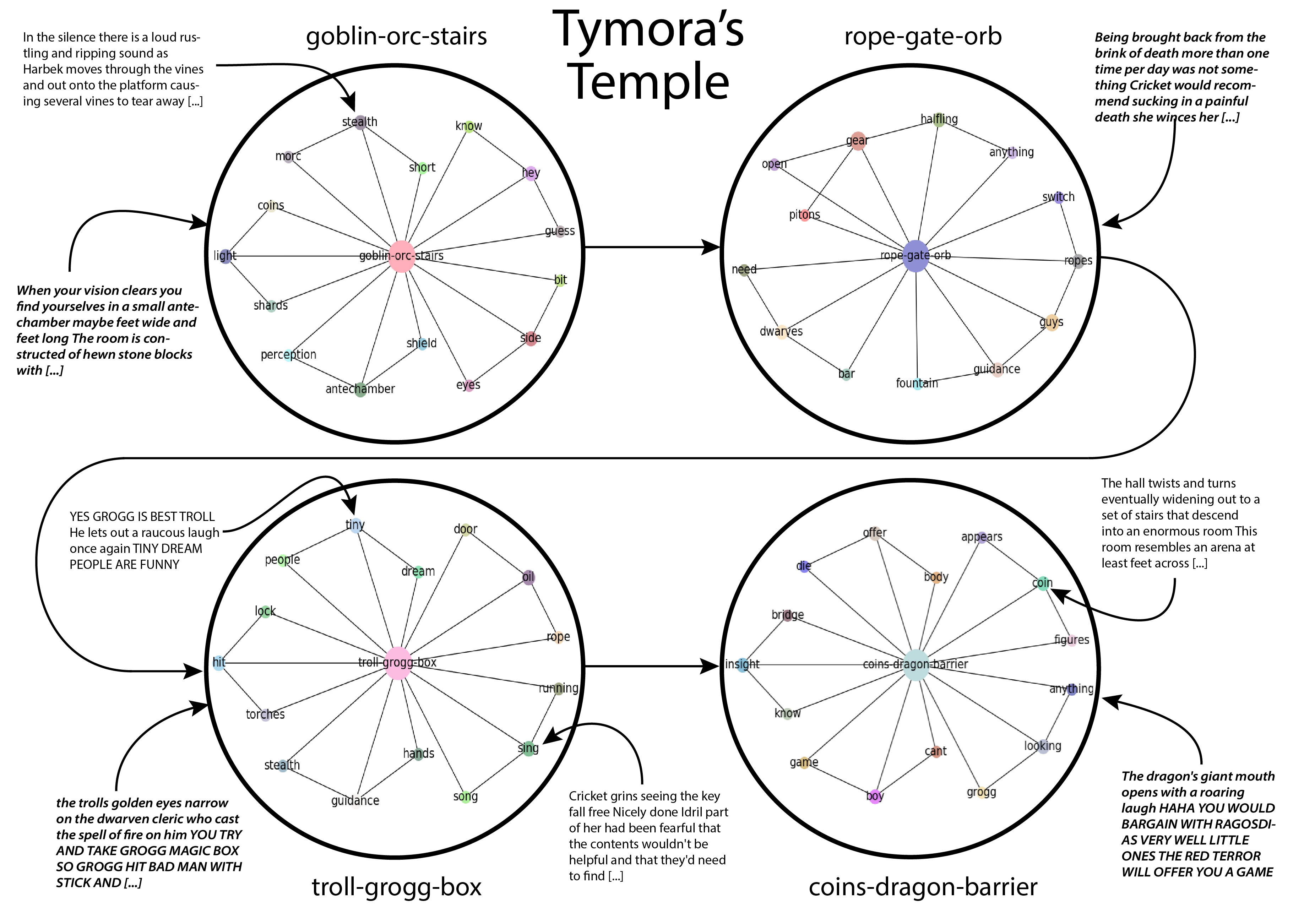}}
			\caption[final map]{Final map\textsuperscript{\footnotemark}} \label{fig:final_map}
		\end{figure}
		\footnotetext{Full size at antibubbles.com/tymoramap.png}
	}

	\section{Discussion}
	Maps are a unique class of diagram in that they portray a continuous and unique spatial relationship between all the elements that are contained within the borders of the map. This differs from other types of diagrams such as network graphs, timelines, and charts. This continuous spatial relationship means that a line drawn in any orientation on the map represents a meaningful trajectory in the space that the map portrays. 
	
	Computing maps of belief places and spaces recalls historical efforts, where maps developed from simple sketches to globally usable Mercator maps that could support coordinated actions over vast distances \cite{black2009revolution}. Similarly, finding \textit{mappable} relationships of \textit{human belief} could elevate the awareness of patterns and structures in this previously unseen environment.

	We selected FTRPGs for this proof-of-concept effort because the alignment-driven \enquote{movement} though the game space matched our needs for understanding how to use repeated, overlapping discussions to construct maps of belief environments. In this regard, fantasy role playing games can be regarded as environments that are constructed entirely from belief, even the rooms that feel so real in the retelling.  From this foundation, we hope to extend this work to broader areas. For example, it may be possible to construct large-scale maps of our shared belief spaces using social media posts from sources like Reddit or Twitter.
	
	But within this study, we also found some unexpected results that reveal the power of such collaborative gameplay to produce results that are startlingly rich and deep.
	
	The first discovery is the viability of our \enquote{naive}, baseline approach of determining salient textual features for the overall game population and individual groups using a Sequential Bag of Words approach. Once stop words and self-identifying terms have been stripped away, selecting the top  population terms, then holding out those terms while selecting the top terms for each group provided a recognizable description of the environment and group approaches to the participants. Assembling these descriptions in sequential order allowed us to assemble our proof-of-concept map based only on the textual interactions of the players. This map is a special type of embedding -- one that is based on the narrative paths of multiple human interactions. As such, it reflects more of our relationship with belief, and is more intuitively navigable than embeddings produced by automated means that do not take these issues into consideration.
	
	The second finding was with respect to how players interacted in this long form experience, where they had time to develop group dynamics. We designed our dungeon so that the last room is a version of the Trolley Problem. This canonical ethical dilemma describes a situation where a bystander is confronted with a runaway trolley that will run down a group of people if nothing is done, or kill one person if a nearby lever is pulled~\cite{thomson1976killing}. In our version, A dragon holds hostage a young boy and an old woman and offers a game. The players must choose one mortal to stay and starve, while the other will go free. If no choice is made, the dragon will devour the entire party and keep both the old woman and the boy in starving captivity. The problem is carefully worded to imply that the party must choose the boy or the woman, while also allowing that there might be a possibility that one of the party can take their place.
	
	The trolley problem has experienced a resurgence in recent years as it provides an ethical framework to understand how we should train/program autonomous systems such as self-driving cars. A team from MIT built a website that crowdsourced this problem, incorporating millions of responses from all over the world. The website would present the participant with a series of image-based scenarios, where one had to click on an image to decide if an imaginary character lives or dies. Using this amassed data, they determined a hierarchy of who should die in a collision. At one end of the spectrum are pets, criminals, and the elderly who fare poorly. At the other end are children. At the top of the hierarchy are the vehicle's passengers, that are spared most often in these scenarios~\cite{awad2018moral}. 
	
	Our results are different. In each of the five completed adventures, the participants decided to self-sacrifice. In fact there were often arguments between the character about who should be the one who stays. The following quote is typical:
	
	\begin{displayquote}
		\textit{\enquote{Get him home.  And deliver my cut of earnings to the people of Phandalin near Neverwinter, my home}. With this, before anyone can stop him, Edmund turns to the dragon. \enquote{I make a counter offer.  In exchange for them (motions to the two caged people). I offer myself to take their place.  I will remain.  I will starve.  You will lose two peasants, and in return you will gain all that I have to offer.  Edmund of house DeVir of Neverwinter.  The last of a noble bloodline of the ruling class.}} (Edmund DeVir: Group 2)
	\end{displayquote}
	
	Why this result? We believe that it might be that the hours invested in the game accesses a \textit{belief space} that is not accessed in the MIT study. A significant feature of the MIT study is that it is designed to elicit a rapid, image-dominant evaluation of the scenario. This affords a superficial consideration of the problem, where the user is not invested in the scenario. In opposition to this approach, the belief space of the dungeon is reachable through the co-creation of textual narratives over time. The players do not just evaluate the problem, then \textit{inhabit} it. These different interfaces enable the navigation of alternate belief spaces. Being able to show these different trajectories through the common salient elements of belief places such as the trolley problem is what we hope to be able to achieve with these types of maps at scale.

	\section{Future Work}
	This paper presents our initial results using a naive text analysis and mapping process. We deliberately made this test dungeon as simple we could and still expect meaningful results. Immediate next steps include validating the map by running an additional adventure where the participants will be provided with the map shown in figure \ref{fig:final_map}, and creating a more sophisticated environments, such as dungeons with a large number of connections, and unconstrained travel within limited spaces (e.g. islands). These progressively more complex environments move us closer to a process that supports the development of large-scale belief networks using large scale, less characterized, social data.

	Now that we have a baseline analytic, we also intend to look at alternate mechanisms for extracting place and space terms/topics, such as using graph convolutional networks to create embeddings~\cite{kipf2016graph}. We are currently experimenting with term embedding~\cite{collobert2011natural}, reading in the entire corpora to produce "grounded" embeddings. We believe that there is considerable promise in exploiting the large language models and embeddings now being developed using transformer machine learning approaches, such as ELMo~\cite{peters2017semi} and BERT~\cite{devlin2018bert}.
	
	We are also looking at mechanisms to scale the approach by automating the role of the dungeon master. Recent work on generating contextual, grounded dialog in fantasy text adventure games~\cite{urbanek2019learning} indicates that this could be a viable option for integrating the mechanics of DM \textit{agents} into these models. This would not only support larger scale studies in the real world, but would create more sophisticated synthetic agents that can be used for more complex models of belief-space navigation in simulation.

	\section{Conclusions}
	
	In this paper, we have described a framework that supports the creation of maps of belief space that can be shown to be correct when compared against the ground truth encoded in the structure of a D\&D adventure. Five online games were run involving 23 participants over a period of several months. Using the processes set up in this study, the process of map creation, from downloading the data to visualizing results are largely automated.
	
	Convergence of terms using time-segmented corpora and straightforward word-counting mechanisms partitioned by user, group and sequence turned out to be surprisingly effective, resulting in stable topic names in less than five runs. An additional benefit was that the process could be described clearly to the participants, who found that the clear explanation added to the credibility of the map. Lastly, the speed and stability of the technique opens up the intriguing possibility that gamespace maps could be generated in real-time for any online environment that produces significant amounts of discussion and alignment within groups.  
	
	The resultant maps were evaluated by the users who participated in the dungeon adventures, and found by significant majorities to meet the requirements of the study - that is, allow users to see themselves in the map, to see the relationship between items and plot a course using the map, and effectively communicate the structures and relationships using the map. 
	
	The current straightforward analytics produces a reasonable lower bound that we can compare against more sophisticated techniques as they are developed. This result implies that the approach is scalable, and can be used for larger, less well-characterized data produced daily in online communities.
	
	Building maps of belief places and spaces can help to expose the relationships between subjective social constructs. In the same way that the text in ship's rutters might be used to create a navigable map of the world, the text of human social interaction can be used to build maps of these unobservable, social spaces. Because maps support individual reasoning about a space, one can find a starting point, choose a destination, and figure out a path to get there. Like with early navigation, the introduction of maps to social domains may transform our understanding of beliefs from constrained and delimited to  panoramic.

	\section{Acknowledgments}
	This effort would not have been possible without the 23 players who spent hours in our dungeon\footnote{antibubbles.com/tymora-thanks.html}. Extra thanks go to Emily Smith and Linda Gregory for participating not only as players, but also as Dungeon Masters.



\begin{thebibliography}{10}
	
	\bibitem{awad2018moral}
	{\sc Awad, E., Dsouza, S., Kim, R., Schulz, J., Henrich, J., Shariff, A.,
		Bonnefon, J.-F., and Rahwan, I.}
	\newblock The moral machine experiment.
	\newblock {\em Nature 563}, 7729 (2018), 59.
	
	\bibitem{barreteau2001role}
	{\sc Barreteau, O., Bousquet, F., Attonaty, J.-M., et~al.}
	\newblock Role-playing games for opening the black box of multi-agent systems:
	method and lessons of its application to senegal river valley irrigated
	systems.
	\newblock {\em Journal of artificial societies and social simulation 4}, 2
	(2001), 5.
	
	\bibitem{belz2013spontaneous}
	{\sc Belz, M., Pyritz, L.~W., and Boos, M.}
	\newblock Spontaneous flocking in human groups.
	\newblock {\em Behavioural processes 92\/} (2013), 6--14.
	
	\bibitem{bikhchandani1992theory}
	{\sc Bikhchandani, S., Hirshleifer, D., and Welch, I.}
	\newblock A theory of fads, fashion, custom, and cultural change as
	informational cascades.
	\newblock {\em Journal of political Economy 100}, 5 (1992), 992--1026.
	
	\bibitem{black2009revolution}
	{\sc Black, J.}
	\newblock A revolution in military cartography?: Europe 1650-1815.
	\newblock {\em The Journal of Military History 73}, 1 (2009), 49--68.
	
	\bibitem{borgstrom2005structure}
	{\sc Borgstrom, R.}
	\newblock Second person: Role-playing and story in games and playable media.
	\newblock In {\em Structure and Meaning in Roleplaying Game Design}. The MIT
	Press, 2010.
	
	\bibitem{collobert2011natural}
	{\sc Collobert, R., Weston, J., Bottou, L., Karlen, M., Kavukcuoglu, K., and
		Kuksa, P.}
	\newblock Natural language processing (almost) from scratch.
	\newblock {\em Journal of machine learning research 12}, Aug (2011),
	2493--2537.
	
	\bibitem{cucker2007emergent}
	{\sc Cucker, F., and Smale, S.}
	\newblock Emergent behavior in flocks.
	\newblock {\em IEEE Transactions on automatic control 52}, 5 (2007), 852--862.
	
	\bibitem{danchin2004public}
	{\sc Danchin, {\'E}., Giraldeau, L.-A., Valone, T.~J., and Wagner, R.~H.}
	\newblock Public information: from nosy neighbors to cultural evolution.
	\newblock {\em Science 305}, 5683 (2004), 487--491.
	
	\bibitem{dant2019dungeons}
	{\sc Dant, A., Feldman, P., and Lutters, W.}
	\newblock Dungeons for science: Mapping belief places and spaces.
	\newblock {\em arXiv preprint arXiv:1904.05216\/} (2019).
	
	\bibitem{deneubourg1989collective}
	{\sc Deneubourg, J.-L., and Goss, S.}
	\newblock Collective patterns and decision-making.
	\newblock {\em Ethology Ecology \& Evolution 1}, 4 (1989), 295--311.
	
	\bibitem{devlin2018bert}
	{\sc Devlin, J., Chang, M.-W., Lee, K., and Toutanova, K.}
	\newblock Bert: Pre-training of deep bidirectional transformers for language
	understanding.
	\newblock {\em arXiv preprint arXiv:1810.04805\/} (2018).
	
	\bibitem{dillman2014internet}
	{\sc Dillman, D.~A., Smyth, J.~D., and Christian, L.~M.}
	\newblock {\em Internet, phone, mail, and mixed-mode surveys: the tailored
		design method}.
	\newblock John Wiley \& Sons, 2014.
	
	\bibitem{epstein2015search}
	{\sc Epstein, R., and Robertson, R.~E.}
	\newblock The search engine manipulation effect (seme) and its possible impact
	on the outcomes of elections.
	\newblock {\em Proceedings of the National Academy of Sciences 112}, 33 (2015),
	E4512--E4521.
	
	\bibitem{ewalt2014dice}
	{\sc Ewalt, D.~M.}
	\newblock {\em Of dice and men: the story of Dungeons \& Dragons and the people
		who play it}.
	\newblock Simon and Schuster, 2014.
	
	\bibitem{fathulla2007diagram}
	{\sc Fathulla, K., and Basden, A.}
	\newblock What is a diagram?
	\newblock In {\em 2007 11th International Conference Information Visualization
		(IV'07)\/} (2007), IEEE, pp.~951--956.
	
	\bibitem{feldman2018maps}
	{\sc Feldman, P.}
	\newblock With maps and mobs: Searching for trustworthiness using belief
	spaces.
	\newblock In {\em Proceedings of the 2018 Conference on Human Information
		Interaction \& Retrieval\/} (2018), CHIIR '18, pp.~351--353.
	
	\bibitem{feldman2018one}
	{\sc Feldman, P., Dant, A., and Lutters, W.}
	\newblock This one simple trick disrupts digital communities.
	\newblock In {\em 2018 IEEE 12th International Conference on Self-Adaptive and
		Self-Organizing Systems (SASO)\/} (2018), IEEE, pp.~50--59.
	
	\bibitem{finley1978homer}
	{\sc Finley, J.~H.}
	\newblock {\em Homer's Odyssey}.
	\newblock Harvard University Press Cambridge, 1978.
	
	\bibitem{flaxman2016filter}
	{\sc Flaxman, S., Goel, S., and Rao, J.~M.}
	\newblock Filter bubbles, echo chambers, and online news consumption.
	\newblock {\em Public Opinion Quarterly 80}, S1 (2016), 298--320.
	
	\bibitem{foucault1986other}
	{\sc Foucault, M., and Miskowiec, J.}
	\newblock Of other spaces.
	\newblock {\em diacritics 16}, 1 (1986), 22--27.
	
	\bibitem{goldie2015early}
	{\sc Goldie, M.~B.}
	\newblock An early english rutter: The sea and spatial hermeneutics in the
	fourteenth and fifteenth centuries.
	\newblock {\em Speculum 90}, 3 (2015), 701--727.
	
	\bibitem{gygax1974dungeons}
	{\sc Gygax, G., and Arneson, D.}
	\newblock {\em Dungeons and dragons}, vol.~19.
	\newblock Tactical Studies Rules Lake Geneva, WI, 1974.
	
	\bibitem{harrigan2010second}
	{\sc Harrigan, P., and Wardrip-Fruin, N.}
	\newblock {\em Second person: Role-playing and story in games and playable
		media}.
	\newblock The MIT Press, 2010.
	
	\bibitem{hegselmann2002opinion}
	{\sc Hegselmann, R., Krause, U., et~al.}
	\newblock Opinion dynamics and bounded confidence models, analysis, and
	simulation.
	\newblock {\em Journal of artificial societies and social simulation 5}, 3
	(2002).
	
	\bibitem{heinsoo2008dungeons}
	{\sc Heinsoo, R., Collins, A., and Wyatt, J.}
	\newblock {\em Dungeons \& Dragons Player's Handbook: Arcane, Divine, and
		Martial Heroes: Roleplaying Game Core Rules}.
	\newblock Wizards of the Coast, 2008.
	
	\bibitem{isozaki1992mechanism}
	{\sc Isozaki, H., and Shoham, Y.}
	\newblock A mechanism for reasoning about time and belief.
	\newblock In {\em Fifth Generation Computer Systems\/} (1992), pp.~694--701.
	
	\bibitem{jacob2006sovereign}
	{\sc Jacob, C., et~al.}
	\newblock {\em The sovereign map: Theoretical approaches in cartography
		throughout history}.
	\newblock University of Chicago Press, 2006.
	
	\bibitem{kipf2016graph}
	{\sc Kipf, T.~N., and Welling, M.}
	\newblock Semi-supervised classification with graph convolutional networks.
	\newblock {\em Computing Research Repository (CoRR) abs/1609.02907\/} (2016).
	
	\bibitem{korzybski1958science}
	{\sc Korzybski, A.}
	\newblock {\em Science and sanity: An introduction to non-Aristotelian systems
		and general semantics}.
	\newblock Institute of GS, 1958.
	
	\bibitem{kosala2000web}
	{\sc Kosala, R., and Blockeel, H.}
	\newblock Web mining research: A survey.
	\newblock {\em SIGKDD Explor. Newsl. 2}, 1 (June 2000), 1--15.
	
	\bibitem{kritikopoulos2007wordrank}
	{\sc Kritikopoulos, A., Sideri, M., and Varlamis, I.}
	\newblock Wordrank: A method for ranking web pages based on content similarity.
	\newblock In {\em 24th British National Conference on Databases (BNCOD'07)\/}
	(2007), IEEE, pp.~92--100.
	
	\bibitem{le1897crowd}
	{\sc Le~Bon, G.}
	\newblock {\em The crowd: A study of the popular mind}.
	\newblock Fischer, 1897.
	
	\bibitem{martindale1992clockwork}
	{\sc Martindale, C.}
	\newblock {\em The clockwork muse: The predictability of artistic change}.
	\newblock Basic Books, 1992.
	
	\bibitem{moscovici1994conflict}
	{\sc Moscovici, S., and Doise, W.}
	\newblock {\em Conflict and consensus: A general theory of collective
		decisions}.
	\newblock Sage, 1994.
	
	\bibitem{olfati2006flocking}
	{\sc Olfati-Saber, R.}
	\newblock Flocking for multi-agent dynamic systems: Algorithms and theory.
	\newblock {\em IEEE Transactions on automatic control 51}, 3 (2006), 401--420.
	
	\bibitem{olfati2007consensus}
	{\sc Olfati-Saber, R., Fax, J.~A., and Murray, R.~M.}
	\newblock Consensus and cooperation in networked multi-agent systems.
	\newblock {\em Proceedings of the IEEE 95}, 1 (2007), 215--233.
	
	\bibitem{parkinson2018similar}
	{\sc Parkinson, C., Kleinbaum, A.~M., and Wheatley, T.}
	\newblock Similar neural responses predict friendship.
	\newblock {\em Nature communications 9}, 1 (2018), 332.
	
	\bibitem{peters2017semi}
	{\sc Peters, M.~E., Ammar, W., Bhagavatula, C., and Power, R.}
	\newblock Semi-supervised sequence tagging with bidirectional language models.
	\newblock {\em arXiv preprint arXiv:1705.00108\/} (2017).
	
	\bibitem{pirolli1999information}
	{\sc Pirolli, P., and Card, S.}
	\newblock Information foraging.
	\newblock {\em Psychological review 106}, 4 (1999), 643.
	
	\bibitem{reynolds1987flocks}
	{\sc Reynolds, C.~W.}
	\newblock Flocks, herds and schools: A distributed behavioral model.
	\newblock In {\em Proceedings of the 14th Annual Conference on Computer
		Graphics and Interactive Techniques\/} (1987), SIGGRAPH '87, ACM, pp.~25--34.
	
	\bibitem{rohl2008mapping}
	{\sc R{\"o}hl, T., and Herbrik, R.}
	\newblock Mapping the imaginary-maps in fantasy role-playing games.
	\newblock In {\em Forum Qualitative Sozialforschung\/} (2008), vol.~9.
	
	\bibitem{salganik2008leading}
	{\sc Salganik, M.~J., and Watts, D.~J.}
	\newblock Leading the herd astray: An experimental study of self-fulfilling
	prophecies in an artificial cultural market.
	\newblock {\em Social psychology quarterly 71}, 4 (2008), 338--355.
	
	\bibitem{Samory2017Quotes}
	{\sc Samory, M., Cappelleri, V.-M., and Peserico, E.}
	\newblock Quotes reveal community structure and interaction dynamics.
	\newblock In {\em Proceedings of the 2017 ACM Conference on Computer Supported
		Cooperative Work and Social Computing\/} (New York, NY, USA, 2017), CSCW '17,
	ACM, pp.~322--335.
	
	\bibitem{simon1996sciences}
	{\sc Simon, H.~A.}
	\newblock {\em The sciences of the artificial}.
	\newblock MIT press, 1968.
	
	\bibitem{thomson1976killing}
	{\sc Thomson, J.~J.}
	\newblock Killing, letting die, and the trolley problem.
	\newblock {\em The Monist 59}, 2 (1976), 204--217.
	
	\bibitem{urbanek2019learning}
	{\sc Urbanek, J., Fan, A., Karamcheti, S., Jain, S., Humeau, S., Dinan, E.,
		Rockt{\"a}schel, T., Kiela, D., Szlam, A., and Weston, J.}
	\newblock Learning to speak and act in a fantasy text adventure game.
	\newblock {\em arXiv preprint arXiv:1903.03094\/} (2019).
	
	\bibitem{yao2018dynamic}
	{\sc Yao, Z., Sun, Y., Ding, W., Rao, N., and Xiong, H.}
	\newblock Dynamic word embeddings for evolving semantic discovery.
	\newblock In {\em Proceedings of the Eleventh ACM International Conference on
		Web Search and Data Mining\/} (2018), ACM, pp.~673--681.
	
	\bibitem{yeshurun2017same}
	{\sc Yeshurun, Y., Swanson, S., Simony, E., Chen, J., Lazaridi, C., Honey,
		C.~J., and Hasson, U.}
	\newblock Same story, different story: the neural representation of
	interpretive frameworks.
	\newblock {\em Psychological science 28}, 3 (2017), 307--319.
	
\end{thebibliography}
\end{document}